\documentclass[twocolumn,showpacs,preprintnumbers,amsmath,amssymb]{revtex4}

\usepackage{graphicx}

\begin{document}

\preprint{TQO-ITP-TUD/01-2011}

\title{Canonical ensemble of an interacting Bose gas:\\
stochastic matter fields and their coherence}
\author{Sigmund Heller}
\affiliation{Institut f\"{u}r Theoretische Physik, 
Technische Universit\"at Dresden, D-01062 Dresden, Germany}
\author{Walter T. Strunz}
\affiliation{Institut f\"{u}r Theoretische Physik, 
Technische Universit\"at Dresden, D-01062 Dresden, Germany}

\date{\today}

\begin{abstract}
We present a novel quantum stochastic evolution equation for a 
matter field describing the canonical state of a
weakly interacting ultracold Bose gas. In the ideal gas limit
our approach is exact. This numerically very stable 
equation suppresses high-energy fluctuations exponentially, which enables
us to describe condensed and thermal atoms within the same formalism. 
We present applications to ground state occupation and fluctuations, 
density profile of ground state and thermal cloud, and ground state number 
statistics. Our main aim are spatial coherence properties which we
investigate through the determination of interference contrast and 
spatial density correlations. Parameters are taken from actual 
experiments \cite{Hof08}.

\end{abstract}

\maketitle

Equilibrium fluctuations in ultracold gases reveal detailed
information about states and phases of interacting
many-body quantum systems \cite{Gri09}. Recent experiments permit to control ultracold 
quantum gases in a hitherto unknown 
precision and to investigate temperature dependent quantities 
like the thermal density and ground state occupancy \cite{Mep10}, spatial correlation functions \cite{Blo00,Foe05,Rit07,Man10} 
density fluctuations \cite{Est06} or interference contrast \cite{Hof08}. 


In this work we determine the canonical state of an interacting ultracold 
Bose gas. A novel quantum stochastic evolution 
equation for a c-number field $\psi(x)$ is presented such that 
canonical quantum 
statistical expectation values can be replaced by an ensemble mean over 
these stochastic matter fields. The equation allows to determine 
coherence properties and other relevant 
observables; it is based on a mean-field type
approximation and strictly valid for the non-interacting case.

Most theoretical descriptions of interacting ultracold Bose gases at finite temperature
are based on grand canonical statistics \cite{Pro08,Gri09}. For actual experiments
involving a fixed and finite number of particles, however, a canonical 
description is natural.
In studying the role of the chosen ensemble, attention so far has been paid to ground
state number fluctuations \cite{Idz99,Koc00}.
While for the ideal gas canonical and grand canonical ensemble give vastly different
predictions \cite{Zif77}, this ceases to be true for interacting gases
in the thermodynamical limit \cite{Gio98}. Our work is based on canonical statistics
right from the start and allows us to not
only investigate occupation fluctuations but also spatial
coherence properties.

Many stochastic field methods exist for the description of Bose gases at finite temperature. All of these are based
on grand canonical statistics, and nicely overviewed and compared in \cite{Pro08,Cock10}. 
In the truncated Wigner approach, the evolution of the field Wigner functional 
is determined approximately from a sampling over random initial fields whose
dynamics is given by the Gross-Pitaevskii equation  \cite{Ste98, Sin02}. 
Based on a quantum kinetic theory and a separation of condensed and 
non-condensed part, Gardiner and co-workers derive a stochastic Gross-Pitaevskii equation \cite{Gar02}. With a similar result, a functional
integral approach to the evolution of the field Wigner distribution is worked 
out in an approach by Stoof and co-workers \cite{Sto97}. Care has to be 
taken with respect to the white noise driving these equations.
Exact methods based on the positive P-representation are used by Drummond 
and co-workers \cite{Khe04}. It is possible to use this approach for 3D 
systems; still, the long-time numerical solution has to be exercised with
caution.
We see the strength of our approach in
its unified applicability to a vast number of different phenomena: from
ground state fluctuations to properties of the thermal cloud, to
coherence properties and contrast
in Bose gas interferometry. For the latter we obtain nice agreement 
with experiments of the Schmiedmayer group that may be 
well described by Luttinger liquid theory \cite{Hof08} or by a stochastic phase model \cite{Sti10}. 

Two properties of our novel equation should be emphasized: first, 
unlike in our previous attempt \cite{hellerstrunz01}, the equation is
{\it not} norm preserving. Still, the norm fluctuations are small
compared to those of related stochastic equations used for grand 
canonical simulations.
Secondly, as in \cite{hellerstrunz01}, ultraviolet cutoff problems 
do not appear due to the use of the Glauber-Sudarshan $P$-function:
effectively, our treatment leads to spatially correlated noise, unphysically
large momentum kicks are suppressed.
These properties afford a very stable numerical solution of full 3D problems,
using arbitrary trap potentials. 
Due to lack of space
and its current interest, we here concentrate on 1D gas interference. Still, we
emphasize that we are also able to treat full 3D gases within our approach
\cite{heller_strunz_3D}.

We propose to use the stochastic (Ito) equation
\begin{eqnarray}\label{quantumequation}
d |\psi\rangle&=&
-\frac{1}{\hbar}\left((\Lambda+i)H-\Lambda\frac{N}{\langle\psi|\psi\rangle}
H  e^{- H/kT}\right)|\psi\rangle dt
\nonumber\\&&+ \sqrt{\frac{2\Lambda}{\hbar}}\sqrt{H e^{- H/kT}}
|d\xi\rangle
\end{eqnarray}
for a c-number matter field $\psi(\vec x,t)=\langle \vec x|\psi(t)\rangle$
to determine all equilibrium properties of a weakly interacting Bose gas 
of $N$ particles 
at arbitrary temperature $T$ ($k$ is Boltzmann's constant). 
Throughout, we will refer to (\ref{quantumequation}) as the
{\it stochastic matter field equation} (SMFE) for finite temperature.
Crucially, the operator $H$ is
the effective (mean-field) one-particle energy operator
\begin{equation}\label{stochastic_meanfield}
H = \frac{{\vec p}^2}{2m} + V ({\vec x}) + 
g(N-1)\frac{|\psi({\vec x,t})|^2}{\langle\psi(t)|\psi(t)\rangle}
\end{equation}
such that equ. (\ref{quantumequation}) may also be seen as a stochastic
Gross-Pitaevskii equation.
As usual, $V(\vec x)$ denotes the trap potential, 
the interaction parameter 
$g$ is proportional to the s-wave scattering length $a_s$ and $m$ is the mass 
of a Boson. The parameter $\Lambda$ appearing in (\ref{quantumequation})
is a phenomenological damping rate 
that sets the time scale for transition to equilibrium. 
Its appearance as square root with the fluctuations
reflects a fluctuation-dissipation-relation.
The fluctuating part is driven by complex Ito increments 
$|d\xi\rangle$ with 
$|d\xi\rangle\langle d\xi|=1\!\!1 dt$, 
$|d\xi^\ast\rangle\langle d\xi|=0$.
Note, however that the operator $\sqrt{H e^{- H/kT}}$ acts on 
the noise, effectively leading to spatially 
correlated noise \cite{hellerstrunz01}. 

Before we show the versatility and accuracy of the SMFE in 
applications later, let us sketch
how we arrive at (\ref{quantumequation}). Our aim is to determine
mean values $\langle\ldots\rangle_N = $tr$[\ldots\rho_N]$
with the canonical density operator
\begin{equation}\label{density}
\hat{\rho}_N=\frac{1}{Z_N}e^{-\hat{H}/kT}\hat{\Pi}_N
\end{equation}
in second quantization with Hamiltonian $\hat{H}$,
canonical partition function $Z_N$, 
and projector 
$\hat{\Pi}_N=\sum\limits_{\sum n_k=N}|\{n_k\}\rangle\langle\{ n_k \}|$ 
onto the $N$-particle subspace. 
Normally-ordered matter field correlation functions are expressed
in terms of functional phase space integrals \cite{hellerstrunz01}, 
for instance
\begin{equation}\label{firstcorrelations}
\langle \hat{\psi}^{\dagger}({\vec x})\hat{\psi}({\vec x}')\rangle_N =
\frac{1}{C_N}\int {\mathcal D}[\psi]\;\psi^{\ast}({\vec x})\psi({\vec x}')\, 
W_{N-1}(\psi),
\end{equation} 
with the weight functionals
$W_N(\psi)=\frac{1}{N!}\,\langle \psi|\psi\rangle^{N}\,
e^{-\langle\psi|\psi\rangle}\,P(\psi)$, where
$P$ denotes the Glauber-Sudarshan P-function \cite{Sch01}
of state $\frac{e^{-\hat{H}/kT}}{Z}$ and
$C_N=\int {\mathcal D}[\psi]\; W_{N}(\psi)$. Note
that second (or higher) order correlations require the use of
$W_{N-2}$ (or lower index) in expression (\ref{firstcorrelations}), 
while $C_N$ remains. 

For the ideal gas case ($g=0$), we prove that
the SMFE (\ref{quantumequation}) corresponds to
a Fokker-Planck equation \cite{Gar83} whose stationary solution is
just the weight functional
$W_N(\psi)$.
Thus, equilibrium 
expectation values of the canonical ensemble are obtained from
propagating equ.
(\ref{quantumequation}) and averaging. In practice, we use a 
long-time-average over 
a single trajectory $\psi(x,t)$. 

The SMFE (\ref{quantumequation}) is exact for an
ideal gas; interactions can be included with great success:
we propose to use the single equation (\ref{quantumequation}) with 
(\ref{stochastic_meanfield}), containing
the current stochastic mean field energy, to describe
all properties of weakly interacting Bose gases in a unified way.
Indeed, we emphasize that (\ref{quantumequation}) interpolates smoothly 
between the high-temperature limit $T\gg T_c$, when interactions are 
negligible and thus our description is exact anyway. At
the opposite end, when $T\ll T_c$, the SMFE (\ref{quantumequation}) with 
(\ref{stochastic_meanfield}) is just mean field Gross-Pitaevskii theory 
and is again expected to give good results.
Clearly, the fluctuations we describe are
of thermal origin; quantum fluctuations are
taken into account to some extent through the use of
the P-representation. Note also that it is of crucial importance to
use the current, stochastic $\psi(x,t)$ 
in (\ref{stochastic_meanfield}), such that {\it on average}
$\langle H \rangle_{(N-1)} = {\vec p}^2/2m + V(\vec x)+
g\langle\hat{\psi}^{\dagger}(x)\hat{\psi(x)}\rangle_{(N-1)}$.
The quality of choice (\ref{stochastic_meanfield}) was tested
by solving equation (\ref{quantumequation}) for a two mode system
and comparing with numerically exact quantum results over a wide
temperature range. 

\begin{figure}[htbp]
  \centering
    \includegraphics[width=8.0cm]{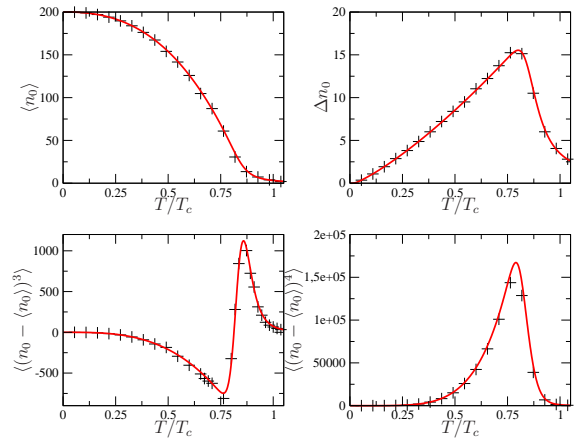}
  \caption{Ground state occupation (top, left), its variance (top, right) $\Delta n_0=\sqrt{\langle (n_0-\langle n_0 \rangle)^2 \rangle}$, third (bottom, left) and fourth (bottom, right) central moments as a function of temperature (scaled with the critical temperature of the thermodynamic limit $T_c=\hbar \omega N^{1/3}/k\zeta(3)^{1/3}$) for an ideal  3D Bose gas of 200 particles in a harmonic trap. The data obtained with the SMFE (black plus signs) is compared with the results of the recursion relation (red solid line).}
  \label{b1}
\end{figure} 

We convince ourselves of the validity of (\ref{quantumequation}) by first
considering an {\it ideal} Bose gas of 200 particles in a 3D harmonic 
trap. In Fig. \ref{b1} results for ground state occupation, its 
variance and further centered 
moments are compared with exact results 
for the canonical ensemble obtained from a recursion relation \cite{Wil97}.

 \begin{figure}[htbp]
  \centering
    \includegraphics[width=8.0cm]{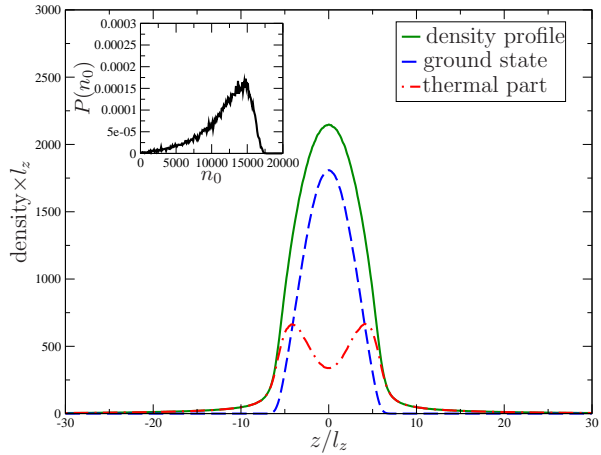}
  \caption{Density profile (green solid line) of an 1D interacting Bose gas of $^{87}$Rb atoms with $\omega_z=2\pi\times 9Hz$, $\omega_{\bot}=2\pi\times 36Hz$ at a temperature of $185nK$ simulated with our SMFE. The ground state contribution (blue dashed line) and the thermal part (red dashed-dotted line) are obtained with the Penrose-Onsager criterion \cite{Pen56} by calculating the full density matrix $\rho(z,z')$. In the upper left corner the ground state number statistics $P(n_0)$ from our simulation is shown.}
  \label{b6}
\end{figure} 

 \begin{figure}[htbp]
  \centering
    \includegraphics[width=8.0cm]{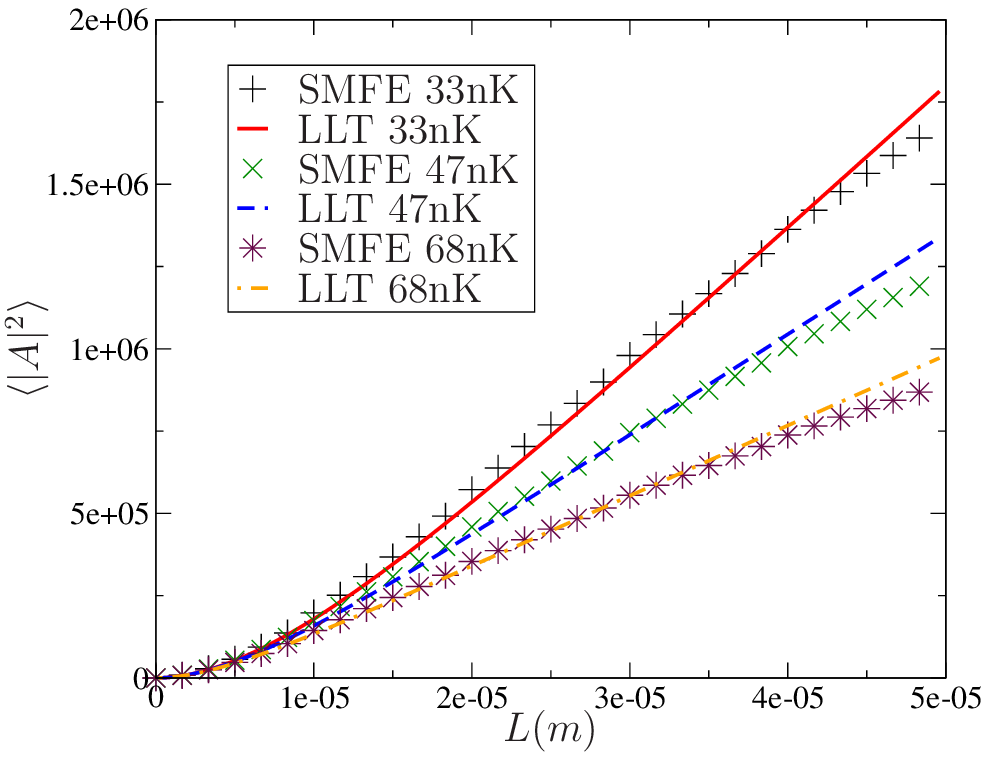}
  \caption{Length dependence of the average contrast $\langle|A(L)|^2\rangle$ of an interference pattern of two uncoupled 1D Bose gases in a harmonic trap for temperatures of $33nK$, $47nK$ and $68nK$ ($n_{1D}\approx50\mu m^{-1}$, $\omega_{\bot}=2\pi\times 3.0kHz$). The data from the SMFE (black plus signs, green crosses, brown stars) is compared to Luttinger-liquid theory (LLT) (red solid line, blue dashed line, yellow dashed-dotted line) which agrees well with the experimental measurements \cite{Hof08}. Small deviations arise from the variation of the density in the harmonic trap, while the Luttinger-liquid theory applies to a uniform density.}
  \label{b3}
\end{figure} 

 \begin{figure}[htbp]
  \centering
    \includegraphics[width=8.0cm]{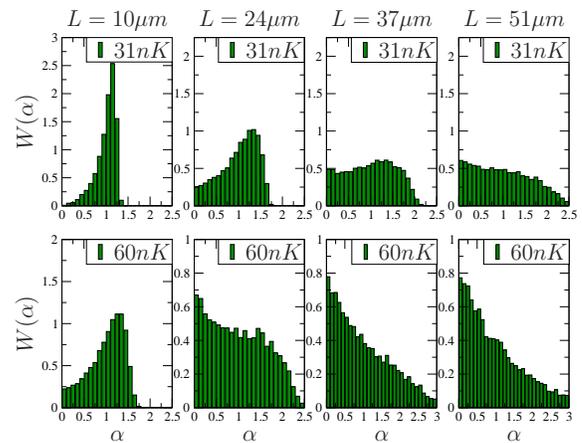}
  \caption{Distribution functions of the interference contrast for different lengths $L$ and different temperatures. The length-dependent normalized interference contrast  $\alpha=\frac{|A|^{2}}{\langle|A|^2\rangle}$ is sampled over $10000$ realizations of the SMFE. The calculation is done for different temperatures and different integration lengths $L$  ($n_{1D}=59\mu m^{-1}$, $\omega_{\bot}=2\pi\times 3.0kHz$).}
  \label{b4}
\end{figure} 

As a first application to the interacting case in Fig. \ref{b6},
the density profile (green solid line) of a $^{87}$Rb quasi-1D gas 
of 20240 atoms in a trap with frequencies $\omega_z=2\pi\times 9Hz$ 
and $\omega_{\bot}=2\pi\times32Hz$ at a temperature of $185 nK$ is shown
(we use 1D coupling constant $g_{1D}=2\hbar\omega_{\bot}a_s$). 
The blue dashed line is the contribution with off-diagonal long range order
(ODLRO) whose wave function $\psi_0(z)$ is obtained from a 
diagonalization of the full $\rho(z,z') = 
\langle\hat\psi^\dagger (z)\hat\psi(z')\rangle_N$
applying the Penrose-Onsager criterion \cite{Pen56}. Moreover, 
a histogram of the stochastic occupation $n_0=\langle\psi_0|\rho|\psi_0\rangle$ 
leads to the ground state number statistics $P(n_0)$ 
(inset in Fig. \ref{b6}).
The average number of particles in state $\psi_0$ turns out to be $12573$.
Our findings are nicely compatible with the ``stochastic Gross-Pitaevskii''
results of \cite{Cock10}, without, however overestimating lowly occupied 
regions.

The SMFE (\ref{quantumequation}) is ideally suited to study 
coherence properties of interacting matter waves through the determination of
spatial correlation functions. As an application we show
results of our SMFE applied to recent experiments in the
Schmiedmayer group \cite{Hof08}:
two independent condensates are prepared in quasi-1D; after expansion they
interfere; the observed interference pattern is integrated over a length
$L$ which determines the contrast $|A(L)|^2$, where
$A(L)=\int\limits_{-L/2}^{L/2}dz\hat{\psi}^{\dagger}_1(z)\hat{\psi}_2(z)$.
Both, mean value
$\langle|A(L)|^2\rangle$ (Fig. \ref{b3}) and the full distribution
$W(\alpha)$ of the normalized moments
defined through
$\int\limits_{0}^{\infty}W(\alpha)\alpha^m d\alpha =
\langle\alpha^m\rangle=
\frac{\langle|A|^{2m}\rangle}{\langle|A|^2\rangle^m}$
are determined (Fig. \ref{b4}).
In \cite{Hof08} it is shown that experimental results are well described
by Luttinger-liquid theory (LLT) to which we will compare our SMFE results.

In Fig. \ref{b3} we show the average contrast
$\langle|A(L)|^2\rangle$ as a function of length $L$ for different
temperatures and find very good agreement with LLT (and thus with experiment).
Deviations for large $L$ arise from density variations along the gas: 
LLT results are based on a uniform density. The gas contains
some $4400$ $^{87}$Rb   
atoms with a central density of 
$n_{1D}\approx 50\mu m^{-1}$. The interaction strength for this 1D case is 
again $g_{1D}=2\hbar\omega_{\bot} a_s$, with $\omega_{\bot}=2\pi\times 3.0kHz$. 
In Fig. \ref{b4} we show the
distribution function of the interference contrast $W(\alpha)$ as
a histogram with 
$\alpha(L)=|A(L)|^2/\langle |A(L)|^2\rangle$ obtained from our simulations. 
We use a central density
$n_{1D}=59\mu m^{-1}$ in line with the experimental setup (see \cite{Hof08}).
LLT predicts a change of the shape of the distribution function with decreasing parameter 
$F=\frac{\hbar^2\pi n_{1D}}{m kT L}$ which is excellently reproduced by 
our simulations.
 \begin{figure}[htbp]
  \centering
    \includegraphics[width=8.0cm]{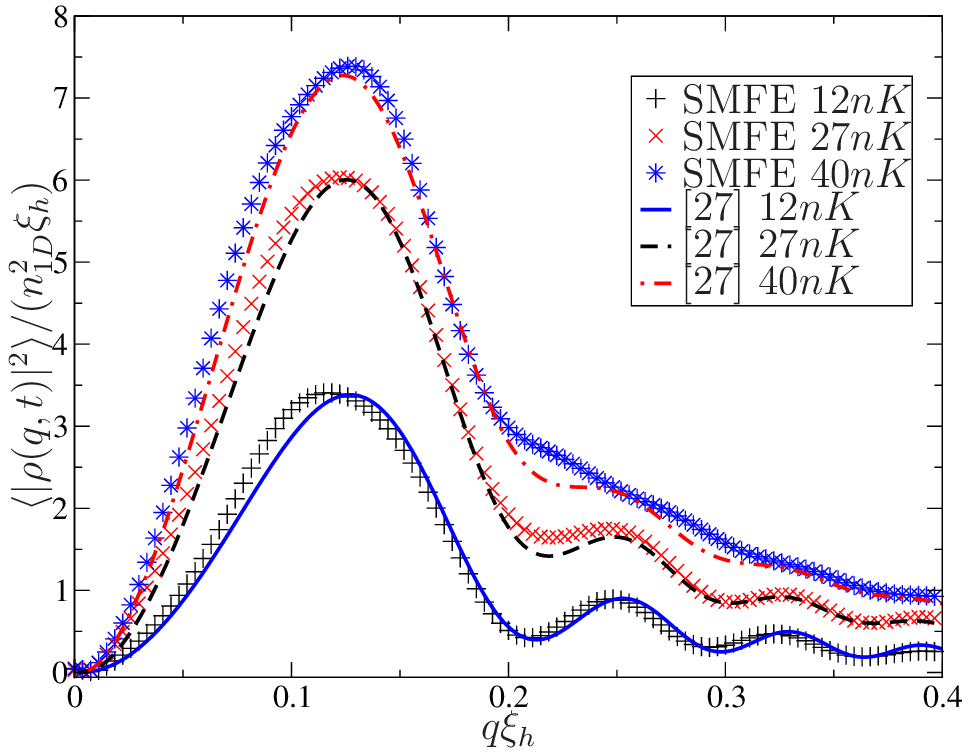}

\caption{Normalized spectrum of density ripples $\langle|\rho(q)|^2\rangle/(n_{1D}^2\xi_h)$ for a weakly interacting Bose gas ($^{87}$Rb) with a central density of $n_{1D}\approx40\mu m^{-1}$, in a trap with transversal frequency $w_{\bot}=2\pi\times 2kHz$;
the expansion time is $t=27ms$, healing length is
$\xi_h=\sqrt{\frac{\hbar}{gm}}$. The simulations of the SMFE are done 
for $12nK$ (black plus signs), $27nK$(red crosses) and $40nK$ (blue stars) 
and compared to theory of \cite{Ima09} ($12nK$ blue solid line, 
$27nK$ black dashed line and $40nK$ red dashed dotted line).}

  \label{b5}
\end{figure}  

Finally, as in \cite{Ima09, Man10},
we investigate two-point density correlation functions after 
expansion 
$g_2(x;t)=\frac{\langle\hat{\psi}^{\dagger}(x;t) \hat{\psi}^{\dagger}(0;t)\hat{\psi}(0;t)\hat{\psi}(x;t)\rangle}{\langle\hat{\psi}^{\dagger}(x,t) \hat{\psi}(x;t)\rangle\langle\hat{\psi}^{\dagger}(0;t)\hat{\psi}(0;t)\rangle}$.
We chose $t=27$ms; note that free (non-interacting) 
expansion can be assumed. 
A gas of about 10000 atoms in a harmonic trap with 
$\omega_{\bot}=2\pi\times2.0kHz$ and a central density of 
$n_{1D}\approx 40\mu m^{-1}$ is used.
As in \cite{Ima09}, we Fourier transform to obtain density ripples 
$\langle|\rho(q)^2|\rangle:=
n_{1D}^2\int\limits_{-\infty}^{\infty}\exp(iqx)[g_2(x;t)-1]$.
In Fig. \ref{b5} our results are compared to calculations from \cite{Ima09}. Again, we see very good agreement.

Let us summarize our achievement: We present a novel quantum stochastic 
matter field equation (SMFE) for a gas of $N$ particles trapped in an arbitrary
potential at any temperature $T$
(canonical ensemble). The equation is strictly valid for a non-interacting 
gas; we include interactions in a stochastic mean-field sense and 
obtain promising results over the entire relevant temperature regime. 
The SMFE is capable of tackling problems in 1D to 3D with arbitrary trapping potentials. 
Results for ground state occupation distribution and density profiles are
shown. Of particular interest is the determination of spatial
correlation functions of arbitrary order. We apply our approach to calculate
interference contrast distributions and density ripples as recently measured.
Our results are in good agreement with Luttinger liquid theory which
has proven to describe the experiments adequately.

We are grateful for inspiring discussions with Antonio Negretti, Carsten Henkel, J\"org Schmiedmayer, Igor Mazets, and Patrick Navez. 
S.~H. acknowledges support by the International Max Planck Research School, 
Dresden. Computing resources have been provided by the 
{\it Zentrum f\"{u}r Informationsdienste und Hochleistungsrechnen} 
(ZIH) at the TU Dresden.

\end{document}